\documentclass[prb,twocolumn,superscriptaddress]{revtex4}
\usepackage{graphicx,bm,amssymb, amsmath}

\begin{document}

\title{Quantum-mechanical wavepacket transport in quantum cascade 
          laser structures}

\date{to appear in Physical Review B 2006}

\author{S.-C. Lee}
\author{F. Banit}
\affiliation{Institut f\"ur Theoretische Physik, Technische Universit\"at
             Berlin, 10623 Berlin, Germany}
\author{M. Woerner}
\affiliation{Max-Born-Institut f\"ur Nichtlineare Optik und  
            Kurzzeitspektroskopie, 12489 Berlin, Germany}
\author{A. Wacker}\email{Andreas.Wacker@fysik.lu.se}
\affiliation{Fysiska Institutionen, Lunds Universitet, Box 118, 22100 Lund, 
            Sweden}
                        
\begin{abstract}
We present a viewpoint of the transport process
in quantum cascade laser structures in which spatial transport of
charge through the structure is a property of 
coherent quantum-mechanical wavefunctions.
In contrast, scattering processes redistribute particles
in energy and momentum but do not directly cause
spatial motion of charge. 

\end{abstract}
                        
\maketitle

\section{Introduction}
Unipolar quantum cascade (QC) laser devices\cite{FaistScience1994} 
are intraband semiconductor 
devices in which the transport processes
and optical (intersubband) transitions which give rise 
to the lasing operation occur only in the conduction band 
of the semiconductor structure. 
This marks a departure from previous interband semiconductor lasers. 
Since the first realization of a QC laser, 
there has been a proliferation of QC laser designs.\cite{Gma01}
QC laser structures are composed of a complicated sequence
of semiconductor layers with different material compositions and
thicknesses. This sequence is repeated tens or hundreds of times
giving rise to a periodic structure in the device. 
Considering the complicated layer composition of these devices, 
the nature of the charge transport process through
these structures is not immediately apparent.

The original concept by Kazarinov and Suris\cite{KazarinovSPS1972} 
was based on coherent tunneling between neighboring wells. It was soon
realized however that scattering plays a crucial role in
establishing the nonequilibrium carrier distribution, in particular
for the depopulation of the lower laser level. 
Thus almost all simulations of 
transport\cite{Don01,Iot01a,BonnoJAP2005,MirceticJAP2005} 
in QC structures assume a
semiclassical model, in which the transport occurs through
scattering transitions between energy eigenstates (Wannier-Stark
hopping \cite{Tsu75}). In such an approach only diagonal elements of the
density matrix (populations or distribution functions) in the 
Wannier-Stark basis are used, and offdiagonal elements 
are neglected (hence, the term semiclassical). 
Ref.~\onlinecite{Iot01a} briefly considered a more fully quantum-mechanical
extension to this semiclassical approach by also including 
offdiagonal elements (coherences) of the density matrix in 
the calculation. This study concluded that quantum-mechanical
coherences were of limited importance to the transport properties.
However, coherent effects were observed experimentally
for the electron transfer from the injector to the upper laser level  
in QC structures.\cite{EIC02a,WOE04} The importance of coherent effects has
also recently been stressed by Ref.~\onlinecite{CallebautJAP2005},
where a simplified calculation scheme is proposed.

The concept that scattering transitions propel the current through
heterostructure systems conflicts with the standard description 
of transport in bulk structures, where complex Bloch 
functions carry the current, while scattering 
redistributes the carriers in momentum space but does not change 
their spatial positions.\cite{AshcroftBOOK1976} We will show that a similar 
description also holds in quantum 
cascade lasers where the current is carried by 
quantum-mechanical wavepackets, and the scattering only
acts locally redistributing carriers in energy or momentum. 

\section{The model}
The quantum-mechanical transport theory we use here is
based on nonequilibrium Green's functions which allow for a
consistent combination of scattering and coherent evolution. 
Quantum-mechanical  coherences are represented by offdiagonal 
elements of the $\bm{G^<}(E)$
correlation function, which is related to the density matrix
$\rho_{\alpha{\bf k},\beta{\bf k}}
=\int dE\, G^<_{\alpha\beta, {\bf k}}(E)/2\pi i$.
The theory is formulated with basis states 
$\Psi_{\alpha{\bf k}}({\bf r}, z) =  
(e^{i{\bf k}\cdot{\bf r}}/\sqrt{\mathcal{A}}) \psi_\alpha (z)$.
Here $\psi_\alpha(z)$ is the envelope function
in the growth direction $z$.
The wavevector ${\bf k}$ and the spatial coordinate
${\bf r}$ are two-dimensional vectors in the plane
of each semiconductor layer (with normalization area $\mathcal{A}$), 
taking fully into account the three-dimensional nature of the 
structure.

We divide the total Hamiltonian as $\hat{H}=\hat{H}_o
+\hat{H}_{\rm scatt}$.
The free  Hamiltonian $\hat{H}_o$ contains the kinetic energy and 
applied voltage, together with the electron-electron  interaction
in a mean field approximation. It is diagonal in {\bf k},
while $\hat{H}_{\rm scatt}$ describes scattering interactions 
$\mathbf{k} \rightarrow \mathbf{k'}$. 
Here we explicitly take into account 
acoustic phonon and longitudinal optical phonon, 
impurity, and interface roughness scattering processes. 
$\hat{H}_{\rm scatt}$ is treated perturbatively using 
self-energies in the self-consistent
Born approximation. For example, for impurity
scattering, the self-energy has the form
\begin{equation}
\Sigma_{\alpha \alpha', {\bf k}}^{<,{\text{imp}}} (E)
= \sum_{\beta\beta',{\bf q}}
\langle V_{\alpha\beta}^{\text{imp}}({\bf q}) 
V_{\beta'\alpha'}^{\text{imp}}(-{\bf q})\rangle
G^{<}_{\beta\beta', {\bf k}-{\bf q}}(E)
\label{eq.ndsig}
\end{equation}
where $V^{\text{imp}}$ represents the impurity scattering potential.
\footnote{For
$\Sigma^{\text{ret}}$ and self-energies 
for other scattering processes, see
Ref.~\onlinecite{Lee02a}. Impurity scattering is treated microscopically
as in Ref.~\onlinecite{Ban05}. We only treat electron-electron 
scattering in a mean field approximation, but it has been
shown that for intersubband transport, electron-impurity
scattering usually dominates over electron-electron 
scattering, see Callebaut, Kumar,
  Williams, Hu, and Reno, Appl. Phys. Lett. {\bf 84}, 645 (2004).}
In the following, we consider both the complete form of  
Eq.~(\ref{eq.ndsig}) with nondiagonal self-energies (ND), as well as 
approximations summarized in Table \ref{tab.pot}.
In each case, the system of equations for the 
Green's functions and self-energies is solved self-consistently, which
is an extensive numerical task. \footnote{We use an iterative scheme 
based on the Broyden method, V.~Eyert, J.  Comp. Physics {\bf 124}, 
271 (1996).}
This results in the determination of the full correlation function
$G^<_{\beta\alpha,{\bf k}}(E)$, which describes the nonequilibrium
state of the device.

Current densities are defined as $J = J_o + J_{\rm scatt}$ with
\begin{equation}
J_o = \frac{ie\langle[\hat{H}_o,\hat{z}]\rangle}{\mathcal{V}\hbar}
    = \frac{2e}{\hbar\mathcal{V}}\sum_{\alpha\beta,{\bf k}}
\int \frac{dE}{2\pi}
[\hat{H}_o, \hat{z}]_{\alpha\beta} G^<_{\beta\alpha,{\bf k}}(E)
\label{eq.jo}
\end{equation}
and $J_{\text{scatt}} = 
ie\langle[\hat{H}_{\text{scatt}},\hat{z}]\rangle/\mathcal{V}\hbar$.
 $e < 0$ is the electron
charge and $\mathcal{V}$ is the sample volume.
In Ref.~\onlinecite{Lee02a} it was shown that 
$J_{\text{scatt}}$ provides the hopping current due to scattering
transitions between the states if one restricts to diagonal Green
functions and self-energies, i.e. neglects coherences between the
states. Thus this part was
referred to as scattering current, reflecting the (mis-)conception that
current flow occurs by a combination 
of both coherent evolution ($J_o$) and relocation
by scattering transitions. In the following we will show that
$J_{\text{scatt}}$ vanishes if coherences are properly taken into
account, while the
entire current is carried by $J_o$.

\begin{table}
\begin{tabular}{l c c c c}
 & $\langle V_{\alpha\alpha} V_{\alpha\alpha}\rangle\;$ & 
 $\;\langle V_{\alpha\beta} V_{\beta\alpha}\rangle\;$ & 
   $\;\langle V_{\alpha\alpha} V_{\beta\beta}\rangle\;$ &  
   $\;\langle V_{\alpha\beta} V_{\beta'\alpha'}\rangle$\\
 \hline
DG   & $\bullet$  & $\bullet$  &  $-$ & $-$\\
ND   & $\bullet$ & $\bullet$  & $\bullet$  & $\bullet$ \\
NDL & $\bullet$  & $-$  & $\bullet$  & $-$\\
\hline
\end{tabular}
\caption{Scattering potential matrix elements included in 
different self-energy models.
$\bullet$ ($-$) indicates scattering potential matrix elements 
included (excluded) from the self-energy. 
The scattering potential $V$ may represent impurity,
interface roughness, or phonon scattering.
The angle brackets represent an averaging over the impurity 
distribution for impurity scattering, the distribution of thickness
fluctuations for interface roughness scattering, and phonon modes for
phonon scattering. DG indicates diagonal self-energies,
while ND represents nondiagonal self-energies. NDL restricts
the scattering matrix elements to local terms $V_{\alpha\alpha}$ 
for ND. }
\label{tab.pot}
\end{table}

{\em Basis states:} The basis functions $\psi_\alpha(z)$
can be chosen in different ways. In theory, the
choice of basis states should not affect
the physical results. In practice, this choice can influence
the physical description in several ways: i) approximations
are always necessary to perform the theoretical
calculation or to facilitate the numerical computation, and
the interplay between the chosen basis and the approximations
may improve or reduce the validity of the approximations and
hence affect the physical result, ii) different choices of basis
states can reveal different physical aspects of the problem.
For instance, as we show in this
paper, working with position eigenstates casts light on spatial
aspects of the problem. 

In our earlier work,\cite{Lee02a} spatially-localized
Wannier states were used as an orthonormal set of $\psi_\alpha(z)$,
which can be constructed for the infinitely extended QC structure in
a straightforward manner and generate a well-defined periodic Hamiltonian.
A second type of states are the Wannier-Stark states, which are the
eigenstates of $\hat{H}_o$. They can be easily obtained by
diagonalizing $\hat{H}_o$ in the basis of Wannier states. (This
procedure avoids the common use of an artificial  spatial confinement.)
A third type of states constitutes the position 
eigenstate basis, using the eigenfunctions of the
position operator $\hat z$. Again we start with the Wannier basis, 
diagonalize $\hat z$, and transform $\hat {H}_o$ into the new basis.
In all bases we evaluate the scattering matrix elements directly
for the respective basis functions.

{\em Structures:} The data presented below was obtained using two
typical examples:
(A) a midinfrared QC
laser\cite{Sir98}  and 
(B) a THz QC laser.\cite{Wil03} In the calculation
we restrict the number of basis functions to 9 (5) per period
for structures A (B).

\section{Vanishing of $J_{\text{scatt}}$}
In Ref.~\onlinecite{Lee02a}, we neglected the  
offdiagonal elements of the self-energies, 
i.e., only terms with $\alpha = \alpha'$
and $\beta = \beta'$ (diagonal self-energy, DG in Table
\ref{tab.pot}) were included in Eq.~(\ref{eq.ndsig}).
In this restricted theory,
we found $J_o$ and $J_{\text{scatt}}$ to be
similar in magnitude. In contrast, using the more rigorous and 
complete formulation
of the theory reported here, where all self-energy terms
are included (nondiagonal self-energies, ND), 
we find that $J_{\text{scatt}} \approx 0$.
This result is basis invariant
if the complete nondiagonal self-energies (ND) are used, as we
checked this explicitly (numerically) for all three types of basis
sets. Hence the total current density is given by
$J_o$.

The same result has recently been analytically demonstrated
  by Lake and
Pandey:\cite{LakePreprint} In the position eigenstate basis, where 
the matrix $z_{\alpha\beta}$ is diagonal, the
expression (16) for $J_{\text{scatt}}$ of Ref.~\onlinecite{Lee02a}
becomes
\begin{multline*}
J_{\text{scatt}}=\frac{2e}{\hbar{\cal V}}\sum_{\bf k}
\int\frac{dE}{2\pi}\sum_{\alpha\beta}z_{\alpha\alpha}
\Big(
G_{\alpha\beta,{\bf k}}^<\bm{\Sigma}^{\rm adv}_{\beta\alpha,{\bf k}}\\
+G^{\rm ret}_{\alpha\beta,{\bf k}}\bm{\Sigma}^{<}_{\beta\alpha,{\bf k}}
-\Sigma^{<}_{\alpha\beta,{\bf k}}G^{\rm adv}_{\beta\alpha,{\bf k}}
-\Sigma^{\rm ret}_{\alpha\beta,{\bf k}}G^<_{\beta\alpha,{\bf k}}\Big)
\end{multline*}
Now the term in brackets corresponds to the right-hand side of the
continuity equation as given in Eq.~(57) of
Ref.~\onlinecite{LakeJAP1997}, which has to vanish. See also Section 2.4 of
Ref.~\onlinecite{MahanPR1987}.

A second line of argument runs as follows:
For microscopic scattering potentials 
which only depend on $\hat{\mathbf{r}}$ (but not 
on momentum $\hat{p}$), the commutator
$[\hat{H}_{\text{scatt}},\hat{z}]$ vanishes and hence 
$J_{\text{scatt}}=0$  by its definition. Thus $J_{\text{scatt}}=0$ 
holds for a wider class of scattering processes, including 
electron-electron scattering, than studied numerically here.
These analytic arguments demonstrate that the previous 
observation of a finite $J_{\text{scatt}}$ in Ref.~\onlinecite{Lee02a} is an 
unphysical result caused by the DG-approximation in the self-energies 
for a nonlocal basis of Wannier functions.

\section{Spatially-local scattering in $J_o$}
Although $J_{\text{scatt}}$ vanishes this does not imply that
scattering processes are absent from the transport process.
These processes do not appear explicitly in Eq.~(\ref{eq.jo}) 
but they act implicitly in determining $\mathbf{G}^<(E)$, 
and hence in driving the current $J_o$. We now investigate
the role played by scattering processes in determining $J_o$. 
In particular, we seek to establish if each scattering event results
in a spatial displacement of charge, or if a scattering process
acts locally resulting in energy redistribution but with
no accompanying spatial transport.

\begin{figure}
\includegraphics[width=0.95\columnwidth,keepaspectratio]{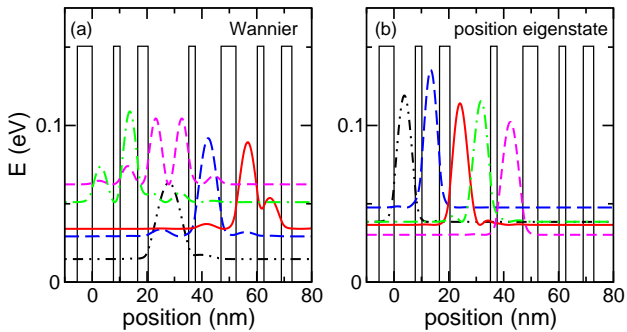}
\caption{(Color online.) Wavefunctions (modulus-squared) for Structure B. 
(a) Wannier, (b) Position eigenstates.}
\label{fig.wfa}
\end{figure}

This issue can be conveniently addressed within the basis of position
eigenstates.
In contrast to the Wannier wavefunctions [Fig.~\ref{fig.wfa}(a)],
these position eigenfunctions are localized to
within a single well layer in the structure 
[see Fig.~\ref{fig.wfa}(b)]
and the scattering potential 
matrix elements $\langle \alpha | V |\beta \rangle$ are 
significantly smaller for $\alpha \neq \beta$ than for $\alpha = \beta$.
We therefore compare the effect of excluding or including 
certain scattering potential terms in the self-energies, 
see Table \ref{tab.pot}. 

\begin{figure}
\includegraphics[width=0.95\columnwidth,keepaspectratio]{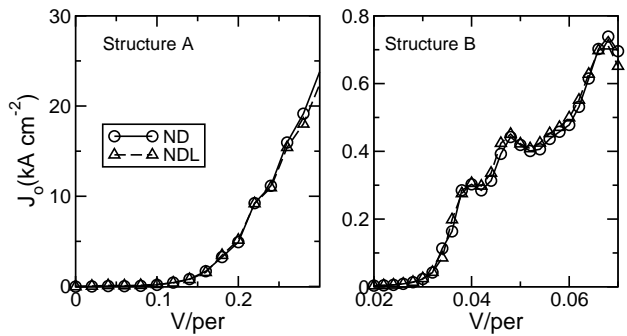}
\caption{Current-voltage characteristic in position 
eigenstate basis with self-energy models ND and NDL. We display $J_0$;
$J_{\rm  scatt}\approx 0$ in both cases.}
\label{fig.jfrei}
\end{figure}

Fig.~\ref{fig.jfrei} shows 
that $J_o^{\rm NDL}$ and $J_o^{\rm ND}$ are almost 
identical for both samples. This indicates that the 
potential terms $\langle V_{\alpha\alpha} V_{\alpha\alpha}\rangle$ 
and $\langle V_{\alpha\alpha} V_{\beta\beta}\rangle$ constitute
the main contribution to the self-energies
in the calculation of $J_o^{\rm ND}$, our 
most rigorous formulation of the
current.  These scattering terms 
involve only purely local (in space) transitions, as
$V_{\alpha\alpha}$ represents a transition matrix element
between states with the same envelope 
function $\psi_\alpha (z) $, i.e.,
localized to the same well. This implies
that the scattering processes act only locally 
in space. They do not cause spatial displacement 
of charge, but act only to redistribute momentum {\bf k} and 
energy.

A key point of our argument was the use of the position
eigenstate basis which enabled us to demonstrate that the
dominant scattering processes in $J_o$ act locally and therefore
do not contribute to the spatial transport of charge by $J_o$.
This result is not evident if we work in another basis, e.g.,
Wannier or Wannier-Stark, where the wavefunctions are
far more delocalized and off-diagonal matrix elements
$V_{\alpha\beta}$ play a significant role. 
Nevertheless, physical 
mechanisms, such as the nature of transport, must be basis invariant.
Indeed, as long as the full self-energies are taken into account, 
the expression for the currents $J_0$ and $J_{\rm  scatt}$ are 
invariant under basis transformations. This demonstrates that the same
result is recovered even in the case of a delocalized basis, where
the interplay of all types of matrix elements in Table  
\ref{tab.pot} reproduces the locality of scattering transitions. 
In contrast, the
neglect of certain matrix element combinations 
(such as the DG-approximation) generates
spurious nonlocal scattering transition in a delocalized basis.

\section{Transport of $J_o$ by coherent wave functions}
As we have just seen, the scattering processes which contribute
to $J_o$  have a purely local effect which does not 
spatially displace carriers.
Hence, we conclude that the transport of charge by $J_o$
through the structure must arise as a property of 
the quantum-mechanical wavefunction. 
We can obtain a more concrete visualization 
of this transport mechanism by resolving $J_o$ in 
energy and space:
\begin{equation}
J_o(E,z) = \frac{e}{\pi \mathcal{A}}\sum_{\alpha\beta,{\bf k}}
{\text{Re}}\left\{\frac{-\hbar}{m(z)}
\psi^*_{\alpha}(z)\frac{\partial\psi_{\beta}(z)}{\partial z}
G^<_{\beta\alpha,{\bf k}}(E)\right\}.
\label{eq.joez}
\end{equation}

\begin{figure}
\includegraphics[width=0.9\columnwidth,keepaspectratio]{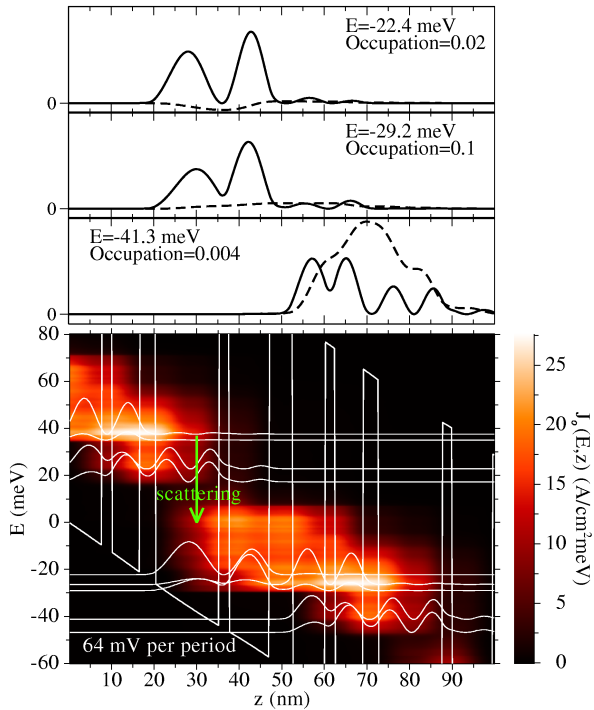}
\caption{(Color online.) {\em Lower panel:}
Spatially and energetically resolved current from Eq.~(\ref{eq.joez})
for sample B.
The wavefunctions depict some quasi-levels,
which are eigenstates of $\hat{H}_o$ at $\mathbf{k} = 0$ 
renormalized by the scattering interactions. {\em Upper panels:} 
Probability density $|\phi_{n{\bf k}}(z)|^2$ (black line)
and current 
$\propto {\text{Re}}\{-i\phi_{n{\bf k}}^*\phi'_{n{\bf k}}(z)/m(z)\}$ 
(dashed line) 
for the complex wave functions $\phi_{n{\bf k}}(z)$ for ${\bf k}=0$.
$n$ corresponds to the largest eigenvalue $f_{n{\bf 0}}(E)$ for the
respective values of $E$.
These energies  are chosen as local maxima 
$E_{\text{max}}$ 
of $f_{n{\bf 0}}(E)$ 
with FWHM $\Delta E$ ($\sim 1$ meV) and the
respective occupations have been approximated by
$(\pi/2)f_{n{\bf 0}}(E_{\text{max}})\Delta E$. }
\label{fig.jez}
\end{figure}

The lower panel of Fig.~\ref{fig.jez} show that 
the current $J_o(E,z)$  flows at different 
energies in different spatial regions of the structures.
To satisfy the equation of continuity, scattering interactions
enable energy transitions which occur locally as discussed above
(an example is depicted by the vertical arrow 
in the figure). Thus, scattering gives rise to a redistribution 
of electrons in energy but does not cause a spatial displacement 
or spatial transport of charge.\footnote{As a further check on the
calculation, current continuity requires that the integration
of Eq.~(\ref{eq.joez}) over energy should result in $J_o(z) = \int dE J_o(z,E)
=\text{a constant}$. Numerically, we observe a spatial variation due to
the limited number of basis states used in the calculation.
Increasing the number of basis states to 8 Wannier states
per period (also used in Fig.~\ref{fig.jez}) reduces this variation
to $< 10$\%.}

\begin{figure}
\includegraphics[width=0.9\columnwidth,keepaspectratio]{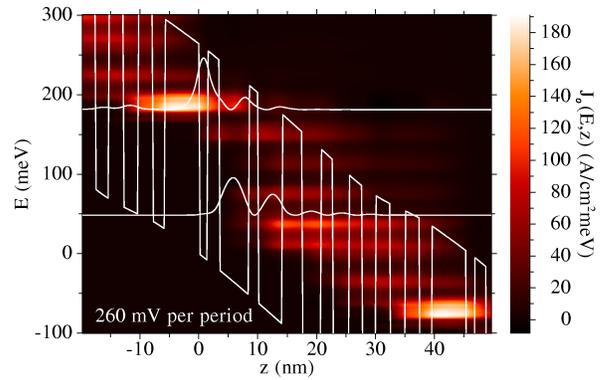}
\caption{(Color online.) 
Spatially and energetically resolved current from Eq.~(\ref{eq.joez})
for sample A. The quasi-levels, which are eigenstates of 
$\hat{H}_o$ at $\mathbf{k} = 0$ renormalized by the scattering
interactions are shown for the upper and
lower laser state for comparison.}
\label{fig.jezSir}
\end{figure}

Fig.~\ref{fig.jezSir} shows the corresponding result for sample A.
Here the current density is restricted to a narrow
energy range while crossing the thick barrier (around $z=-2$ nm) 
between the injector and the active region, 
thus providing the desired selective feeding of the
upper laser level. This demonstrates how this representation allows
for a detailed insight into the microscopic operation of the device.

An especially intuitive form of Eq.~(\ref{eq.joez}) can be obtained
by transforming into an (energy-dependent) basis of eigenstates
of the hermitian matrix $-i/(2\pi)G^<_{\beta\alpha,{\bf k}}(E)$ 
which has real eigenvalues $f_{n{\bf k}}(E)$. While 
$f_{n{\bf k}}(E)$ exhibits sharp peaks in energy, the 
eigenstates $\phi_{n{\bf k}}(z)$
exhibit only a weak energy dependence (not shown for brevity). 
Then we find
\begin{equation}
J_o(E,z) =\frac{2e}{\mathcal{A}}
\sum_{n{\bf k}}f_{n{\bf k}}(E){\rm Re}\left\{
\phi^*_{n{\bf k}}(z)\frac{(-i\hbar)}{m(z)}
\frac{\partial\phi_{n{\bf k}}(z)}{\partial z}
\right\}.
\end{equation}
Thus, the energy-resolved current is carried 
by wave functions $\phi_{n{\bf k}}(z)$ 
with occupation densities $f_{n{\bf k}}(E)$ 
(the factor 2 is due to spin). 
This is analogous to
the Bloch functions and their occupation in standard bulk transport.
The upper panels of Fig.~\ref{fig.jez} show that 
these current carrying states $\phi_{n{\bf k}}(z)$
extend over a range of one period. This length
scale gives a measure of the coherence length, i.e., the distance
over which charge is transported by a single coherent wave packet.
The peaks in density $|\phi_{n{\bf k}}(z)|^2$ 
are localized in the wells as charge tends to accumulate in the
low-potential regions. The currents 
carried by the wave functions $\phi_{n{\bf k}}(z)$
are more evenly distributed which reflects
the ability of the wave packets to transport charge across
the structure.
Note the nonvanishing divergence of 
current density which reflects the local balance of 
scattering processes. Thus, these complex wave functions
$\phi_{n{\bf k}}(z)$ can be viewed as generalized states for a
nonequilibrium system, where the drive by bias and
scattering processes compensate.

The currents associated with 
these wavepackets can differ significantly:
The state at -29.2 meV (corresponding to the upper laser 
level) has only a small current, but a high occupation. 
In contrast, the state at $E=-41.3$ meV (lower laser level) 
exhibits a 
strong current to the right, effectively moving
electrons away from the lasing
transition which occurs vertically between these states.
There are also states exhibiting local current flow to the left, 
e.g., at $-22.4$ meV. 
Such considerations may provide new tools for the
optimization and design of new laser structures.

\begin{figure}
\includegraphics[width=0.95\columnwidth,keepaspectratio]{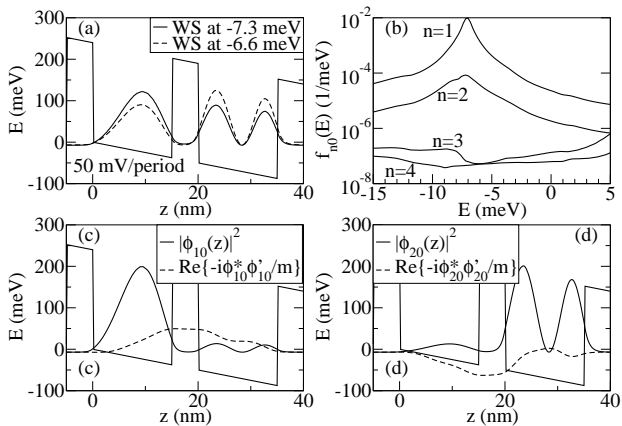}
\caption{States for the superlattice of Ref.~\onlinecite{ZEU96}.
(a) Doublet of Wannier-Stark states. (b) 
Eigenvalues $f_{n{\bf k}}(E)$ of $-i/(2\pi)G^<_{\beta\alpha,{\bf k}}(E)$ 
for ${\bf k}=0$. (c,d) Corresponding eigenfunctions $\phi_{n{\bf
k}}(z)$ at $E=-7\mathrm{meV}$ for the two largest eigenvalues, 
$f_{1{\bf k}}$ and $f_{2{\bf k}}$, respectively.}
\label{FigLevelcross}
\end{figure}

The complex states $\phi_{n{\bf k}}(z)$ are of particular importance
at level crossings. This is demonstrated in
Fig.~\ref{FigLevelcross} for a simple superlattice at the resonance
between the ground state and the first excited state of the
neighboring well. Fig.~\ref{FigLevelcross}(a) depicts the
Wannier-Stark states, which are spread over both wells. 
The first two eigenvalues $f_{n{\bf 0}}(E)$ (for ${\bf k}=0$)
exhibit a clear peak at the level energies
[see Fig.~\ref{FigLevelcross}(b)] 
and the corresponding states, carry particles to the right [state 1,
Fig.~\ref{FigLevelcross}(c)] and to the left [state 2,
Fig.~\ref{FigLevelcross}(d)]. However, the occupation of the state 1 is
about two orders of magnitude larger thus providing a strong current
flux over the barrier.

Our viewpoint of current-carrying coherent wave-functions
differs essentially from the conception that scattering transitions
propel the current.
This manifests itself in two ways: (i)
In a scattering transition picture, the current would stop immediately
once the scattering processes stop ($H_{\text{scatt}}$ drops to zero). 
However, the occupation of current-carrying wave-functions only changes
on the scale of the scattering time, thus the current flow will
continue on this time scale. (ii) The current is sensitive to 
quantum mechanical phases. E.g., phase conjugation would reverse 
the direction of $J_o$, while scattering rates would be unaffected.

\section{Relation to Wannier-Stark hopping model}
In the Wannier-Stark hopping model, the current is carried 
by scattering transitions between Wannier-Stark states and
depends only on scattering rates and the populations of
these states. The coherences, 
i.e., offdiagonal elements $\rho_{\alpha{\bf k},\beta{\bf k}}$
(diagonal in ${\bf k}$
in contrast to scattering transition amplitudes
$\rho_{\alpha{\bf k},\beta{\bf k}'}$)
of the density matrix,
play no role in this model and are neglected.
There are two  peculiarities 
in this picture: (i) The Wannier-Stark
hopping picture of transport through scattering transitions
and our picture of transport by coherent wavepackets each
provide the entire current 
and yet offer such conflicting views of the transport
mechanism.  (ii) If we examine Eq.~(\ref{eq.jo}) 
expressed in the Wannier-Stark basis,
we see that a diagonal density matrix results in
zero total current.\footnote{This can also be seen in 
Eq.~(4) of Ref.~\onlinecite{Cal84} or Eq.~(9) of Ref.~\onlinecite{Lya95}.}
Thus, coherences are central to the transport process, 
even in the Wannier-Stark basis, see also
Ref.~\onlinecite{IottiPRB2005}. This conflicts with the common
Wannier-Stark hopping picture. How do we reconcile these
contradictions? The answer is that the  Wannier-Stark 
hopping model is derived from Eq.~(\ref{eq.jo}) by expressing, 
in a low-order approximation, the offdiagonal elements of the 
density matrix in terms of the diagonal elements.\cite{Cal84,Lya95,Wac02}
Under conditions where such an approximation is applicable 
the Wannier-Stark hopping model gives the same results
as a full quantum transport model 
(see, e.g., Ref.~\onlinecite{Wac02} for superlattices), 
but it should be remembered that 
the Wannier-Stark hopping model uses populations which are
in actual fact an approximation for the coherences. 

Following the concepts of standard quantum optics, one is tempted to
assume that coherences are created by the electric field and
destroyed by scattering.
However, the opposite is true in the quantum treatment of transport 
within the basis of Wannier-Stark states, which are eigenstates of $\hat{H}_o$
including the electrostatic potential. Here coherences are 
induced by the scattering processes
via the non-diagonal self-energies, which are crucial in a consistent
quantum treatment.

\section{Summary} 
We have shown that the entire current through QC 
structures is carried only by complex, 
quantum-mechanical wavefunctions as shown, e.g., in the upper
panel of Fig.~\ref{fig.jez}. This conclusion follows from
two main results:
(i) The scattering current $J_{\text{scatt}}$ vanishes, 
which directly follows from $[\hat{H}_{\text{scatt}},\hat{z}]=0$,
if the scattering potentials couple to the spatial coordinate of
electron states. 
(ii) In the position basis, only the diagonal 
scattering matrix elements $V_{\alpha\alpha}$ are significant
in determining the current $J_o$, and hence the entire
current. Such matrix elements do not induce spatial translation 
of the electron state within the scattering transition. 
Thus, the principal scattering events which transfer momentum 
and/or dissipate energy in the transport process in quantum 
cascade structures occur only through spatially-local 
transitions which do not give rise to propagation of charge.

The locality of scattering transitions 
contrasts with the simple picture of electrons
hopping between energy eigenstates due to scattering events,
which is a common description in heterostructure systems. 
On the other hand, our point of view is  analogous to 
bulk transport, where the scattering term 
in Boltzmann's equation is local in space,
and the entire current is carried  by Bloch states.

We thank R.~Lake for helpful discussion.
This work was supported by Deutsche Forschungsgemeinschaft
and the Swedish Research Council.

%\bibliographystyle{apsrev}
%\bibliography{ref,references}

\end{document}